\documentclass[prd,twocolumn,aps,showpacs]{revtex4}
\def\[{\left\lbrack}
\def\]{\right\rbrack}

\def\({\left(}
\def\){\right)}
\newcommand{\be}{\begin{equation}}
\newcommand{\ee}{\end{equation}}
\newcommand{\ea}{\end{eqnarray}}
\newcommand{\ba}{\begin{eqnarray}}

\begin{document}


\title{Radiation damping, noncommutativity and duality}

\author{E. M. C. Abreu$^a$\footnote{\sf E-mail: evertonabreu@ufrrj.br}, A. C. R. Mendes$^b$\footnote{\sf E-mail: albert@ufv.br},
C. Neves$^c$\footnote{\sf E-mail: clifford.neves@gmail.com} and W. Oliveira$^d$\footnote{\sf E-mail: wilson@fisica.ufjf.br}} 
\affiliation{${}^{a}$Grupo de F\' isica Te\'orica e Matem\'atica F\' isica,
Departamento de F\'{\i}sica, Universidade Federal Rural do Rio de Janeiro\\
BR 465-07, 23890-000, Serop\'edica, Rio de Janeiro, Brazil\\
${}^{b}$Campus Rio Parana\' iba, Universidade Federal de Vi\c{c}osa,\\
BR 354 - km 310, 38810-000, Rio Parana\' iba, Minas Gerais, Brazil\\
${}^{c}$Departamento de Matem\'atica e Computa\c{c}\~ao, Universidade do Estado do Rio de Janeiro\\
Rodovia Presidente Dutra, km 298, 27537-000, Resende, Rio de Janeiro, Brazil\\
${}^{d}$Departamento de F\'{\i}sica,ICE, Universidade Federal de Juiz de Fora,\\
36036-330, Juiz de Fora, MG, Brazil\\
\bigskip
\today\\
{\it {\bf Dedicated to the memory of Prof. Emerson da Silva Guerra}}}

\begin{abstract}
In this work, our main objective is to construct a $N=2$ supersymmetric extension of the nonrelativistic $(2+1)$-dimensional model describing the radiation damping on the noncommutative plane with scalar (electric) and vector (magnetic) interactions by the $N=2$ superfield technique.  We also introduce a dual equivalent action to the radiation damping one using the Noether procedure.
\end{abstract}

\maketitle

\section{Introduction}
\label{intro}

The study of dissipative systems in quantum theory is of strong interest and relevance either for fundamental reasons \cite{uz} and for its practical applications \cite{cl,whb}.  The explicit time dependence of the Lagrangian and Hamiltonian operators introduces a major difficulty to this study since the canonical commutation relations are not preserved by time evolution.  Then different approaches have been used in order to apply the canonical quantization scheme to dissipative systems \cite{mt,bc}.

One of these approaches is to focus on an isolated system composed by the original dissipative system plus a reservoir.  One start from the beginning with a Hamiltonian which describes the system, the bath and the system-bath interaction.  Subsequently, one eliminates the bath variables which give rise to both damping and fluctuations, thus obtaining the reduced density matrix \cite{cl,bc,fv,cl2,cl3}.

Another way to handle the problem of quantum dissipative systems is to double the phase-space dimensions, so as to deal with an effective isolated system composed by the original system plus its time-reversed copy \cite{bm,ft}.  The new degrees of freedom thus introduced may represented by a single equivalent (collective) degree of freedom for the bath, which absorbs the energy dissipated by the system.

The study of the quantum dynamics of an accelerated charge is appropriated to use the indirect representation since it loses the energy, the linear momentum, and the angular momentum carried by the radiation field \cite{hjp}.  The effect of these losses to the motion of charge is known as radiation damping (RD) \cite{hjp}.

The reaction of a classical point charge to its own radiation was first discussed by Lorentz and Abraham more than one hundred years ago, and never stopped being a source of controversy and fascination \cite{becker,lorentz}.  The most disputable aspects of the Abraham-Lorentz theory are the self-acceleration and preacceleration.  Self-acceleration refers to classical solutions where the charge is under acceleration even in the absence of an external field.  Preacceleration means that the charge begins to accelerate before the force is actually applied.

The process of radiation damping is important in many areas of electron accelerator operation \cite{walker}, as in recent experiments with intense-laser relativistic-electron scattering at laser frequencies and field strengths where radiation reaction forces begin to become significant \cite{hk,bula}.

In this paper we introduce a $N=2$ supersymmetric extension  of the radiation damping model completing the $N=1$ supersymmetric version introduced in \cite{mendes,Barone}.  Also a new action dual equivalent to the RD one is obtained using the Noether dualization procedure.  Using the variables introduced in \cite{Plyushchay1} we obtain a new nonvanishing phase space of Poisson brackets.  The Lagrangian is so divided in two parts describing ``external" and ``internal" degrees of freedom in a noncommutative phase space.  In this work, our objective is to describe a $N=2$ supersymmetric extension of the nonrelativistic (2+1)-dimensional model describing the radiation- damping (represented by the equation (\ref{01}) below), on the noncommutative plane introducing an interaction to the free model by the $N=2$ superfield technique.   The introduction of a scalar superpotential for the interaction term permit us to construct $N=2$ supersymmetric Lagrangians. 

This paper is organized in the following distribution: in section 2 we present briefly the $D=2+1$ model and obtain a dual equivalent model through the Noether dualization procedure in section 3.  In section 4  we introduce a symplectic structure in the model in order to introduce the noncommutativity through the variables used in \cite{Plyushchay1}.  In section 5 we promote the supersymmetric extension of the model.  A supersymmetric version through the Hamiltonian formalism is depicted in section 6.  Finally, as usual the conclusions and perspectives are described in the last section. 
 
\section{The model}

In \cite{Lukierski} it was introduced a nonrelativistic classical mechanics free particle model with a Chern-Simons like term
\be\label{01.1}
L=\frac{1}{2}m\dot x_i \dot x_j \,+\,\lambda\varepsilon_{ij} x_i \dot x_j, \;\;i,j=1,2,
\ee
where$\lambda$ has dimension of mass/time, and we can realize the second term as a particular electromagnetic coupling.  To make (\ref{01.1}) quasi-invariant under $D=2$ Galilei symmetry the second term in (\ref{01.1}) was modified and we have that
\be \label{01.2}
L=\frac{1}{2}m\dot x_i \dot x_j -\kappa\varepsilon_{ij}\dot x_i \ddot x_j, \;\;i,j=1,2,
\ee
where $\kappa$ has dimensions of mass $\times$ time.  It can be shown \cite{ls} the quasi-invariance of this Lagrangian just above.
The authors in \cite{Lukierski} demonstrated that the model describes the superposition of a free motion in noncommutative $D=2$ spaces.
A $N=2$ supersymmetric extension of (\ref{01.2}) was accomplished in \cite{Lukierski2} describing particles on the noncommutative space with electric and magnetic interactions.

In \cite{Barone,Albert1} a new approach in the study of radiation damping \cite{hjp} was presented, introducing a Lagrangian formalism to the model in {\it D}= 2+1 dimensions given by
\be\label{01}
L=\frac{1}{2}mg_{ij}\dot x_i \dot x_j -\frac{\gamma}{2}\varepsilon_{ij}\dot x_i \ddot x_j, \;\;i,j=1,2,
\ee
where $\varepsilon_{ij}$ is the Levi-Civita antisymmetric metric, $g_{ij}$ is the pseudo-Euclidean metric given by
\be\label{02}
g=\pmatrix{1 & 0\cr 0 & -1\cr},
\ee
and where, as will be the case throughout the paper, the Einstein convention on the summation of repeated indices is employed. The Lagrangian (\ref{01}) describes, in the hyperbolic plane, the dissipative system of a charge interacting with its own radiation, where the 2-system represents the reservoir or heat bath coupled to the 1-system \cite{mendes,Barone}. The model (\ref{01}) was shown to have the (2+1)-Galilean symmetry and the dynamical group structure associated with that system is the {\it SU}(1,1) \cite{Albert1}. Note that this Lagrangian is similar to the one discussed in \cite{Lukierski} (that is a special nonrelativistic limit of  the particle with torsion investigated in \cite{Plyushchay}), but in this case we have a pseudo-Euclidean metric and the RD constant, $\gamma$, is the coupling constant of a Chern-Simons-like term. It is important to note that, despite the results obtained in this paper being very closely related with the ones from \cite{Lukierski2}, the difference between them is not just the pseudo-Euclidean metric. The physical systems studied are different, where the constant $\gamma$ is not a simple coupling constant, but  depends on the physical properties of the charged particle, like its charge $e$ and mass $m$, being related to the term in its equation of motion which describes an interaction of the charge with its own radiation field.  The radiation-damping constant $\gamma$ make the role of the ``exotic" parameter $\kappa$ in \cite{Lukierski,Lukierski2}.

A supersymmetrized version of the model (\ref{01}) was presented by us in \cite{Albert2} to $N=1$, where we employ a supersymmetric enlargement of the Galileo algebra obtained in \cite{Albert1} and shown that the supersymmetric action can be split into dynamically independent external and internal sectors. 
  
\section{Duality through Dualization}

The bosonization technique that express a theory of interacting fermions in terms of free bosons provides a powerful non-perturbative tool for investigations in different areas of theoretical physics with practical applications \cite{coleman}.  In two-dimensions these ideas have been extended in an interpolating representation of bosons and fermions which clearly reveals the dual equivalence character of these representation \cite{dns}.  In spite of some difficulties, the bosonization program has been extended to higher dimensions \cite{luscher,marino}.

This new technique to perform duality mappings in any dimensions that is alternative to the master action approach has been used in the literature \cite{iw}.  It is based on the traditional idea of a local lifting of a global symmetry and may be realized by an iterative embedding of Noether counter terms.  This technique was originally explored in the context of the soldering formalism \cite{abw,bw} and is explored \cite{iw2,ainrw,binrw} since it seems to be the most appropriate technique for non-Abelian generalization of the dual mapping concept.

There has been a number of papers examining the existence of gauge invariance in systems with second class constraints \cite{vyt}.  Basically this involves disclosing using the language of constraints, hidden gauge symmetries in such systems.  This situation may be of usefulness since one can consider the non-invariant model as the gauge fixed version of a gauge theory.  By doing so it has sometimes been possible to obtain a deeper and more illuminating interpretation of these systems.  Such hidden symmetries may be revealed by a direct construction of a gauge invariant theory out of a non-invariant one \cite{nw}.  The former reverts to the latter under certain gauge fixing conditions.  The advantage in having a gauge theory lies in the fact that the underlying gauge invariant theory allows us to establish a chain of equivalence among different models by choosing different gauge fixing conditions.

As said above, we will use the iterative Noether gauging technique (also called dualization procedure) to achieve this objective.
The important point to stress in this application of the Noether technique is the ability to implement specific symmetries leading to distinct models.  It is an alternative route to establish dual equivalences between gauge and non-gauge theories.  In a nutshell, it is based on the local lifting of the global symmetries present in the non-gauge action.  This is done by iteratively incorporating counter-terms into the action along with a set of auxiliary fields.  Clearly, the resulting embedded theory is dynamically equivalent to the original one \cite{ainrw}.  This alternative approach to the dual transformation is dimensionally independent and sufficiently general to encompass both Abelian and non-Abelian theories.

As the first step, let us rewrite our RD model equation (\ref{01}),
\be
L_0\,=\,{1\over 2}\,m\,g_{ij}\,\dot{x}_i\,\dot{x}_j\,-\,{\gamma\over 2}\,\varepsilon_{ij}\,\dot{x}_i\,\ddot{x}_j,\quad i,j=1,2\;\;,
\ee
hence the variation of this action is
\be
\delta L_0\,=\,J_1^i\,\dot{\eta}_i\,+\,J_2^i\ddot{\eta}_i
\ee
where $\delta x_i=\eta_i$ and the Noether currents are
\ba
J_{1i}&=&m\,g_{ij}\,\dot{x}_j\,-\,{1\over 2}\,\gamma\,\varepsilon_{ij}\ddot{x}_j \\
J_{2i}&=&{1\over 2}\,\gamma\,\varepsilon_{ij}\dot{x}_j \;\;.
\ea
The second step in the iterative method is to construct the action,
\be
L_1\,=\,L_0\,-\,D_1^i\,J_{1i}\,-\,D_2^i\,J_{2i}
\ee
where $D_1^i$ and $D_2^i$ are auxiliary fields which will be eliminated in the process.

The variation of $L_1$ will give us,
\be \label{80}
\delta L_1\,=\,-\,m\,g_{ij}\,D_1^i\,\delta D_1^j\,+\,{1\over 2} \gamma\,\varepsilon_{ij}\,\delta(D_1^i\,D_2^j)
\ee
and the final gauge invariant model is 
\be \label{90}
L_2\,=\,L_1\,+\,{1\over 2}\,m\,g_{ij}\,D_1^i\,D_1^j\,-\,{1\over 2} \gamma\,\varepsilon_{ij}\,D_1^i\,D_2^j
\ee
that automatically compensates for equation (\ref{80}), making $L_2$ gauge invariant and ending the iterative chain.

We have therefore succeed in transforming the global RD theory into a locally invariant gauge theory.  The next step would be to take advantage of the Gaussian character of the auxiliary fields $D_1$ and $D_2$ to rewrite (\ref{90}) as an effective action depending only on the original 
    variable $x_i$ \cite{iw2}.

\section{Noncommutativity}

Introducing a Lagrangian multiplier which equates $\dot x$ to $z$, and substituting all differentiated $x$-variables in the Lagrangian 
(\ref{01}) by $z$-variables, one has a first- order Lagrangian:
\be\label{03}
L^{(0)}=p_i (\dot x_i -z_i ) +\frac{m}{2}g_{ij} z_i z_j -\frac{\gamma}{2}\varepsilon_{ij}z_i \dot z_j,
\ee
which equations of motion can be written, by employing the symplectic structure \cite{Faddeev}, as
\be\label{04}
\omega_{ij}\dot \xi ^j =\frac{\partial H(\xi^j )}{\partial \xi^{i}}
\ee
where the symplectic two form is
\be\label{05}
(\omega)=\pmatrix{{\bf0} & -{\bf 1}_2 & {\bf0}\cr {\bf 1}_2 & {\bf0} & {\bf0} \cr {\bf0} & {\bf0} & -\gamma \varepsilon }
\ee
with
\be\label{06}
{\bf 1}_2 =\pmatrix{1 & 0 \cr 0 & 1}, \;\;\; \varepsilon=\pmatrix{ 0 & 1 \cr -1 & 0 },
\ee
and {\bf 0} denotes the 2$\times$2 null matrix. $H(\xi^l)$ is the Hamiltonian and $\xi^i$ are the symplectic variables.

Now, using the variables introduced in \cite{Plyushchay1} modified as
\be\label{07}
{\cal Q}_i =\gamma(mg_{ij}z_j -p_i ),\; X_i=x_i +\varepsilon_{ij}{\cal Q}_{ij},\; P_i=p_i
\ee
one obtains that
\be\label{08}
L^{(0)} =L^{(0)}_{\rm ext} + L^{(0)}_{\rm int}
\ee
where
\be\label{09}
L^{(0)}_{\rm ext}=P_i \dot X_i + {\gamma\over2}\varepsilon_{ij}P_i \dot P_j -{1\over 2m}g_{ij}P_i P_j ,
\ee
and
\be\label{10}
L^{(0)}_{\rm int}={1\over{2\gamma}}\varepsilon_{ij}{\cal Q}_i \dot{\cal Q}_j +{1\over{2m\gamma^2}}g_{ij}{\cal Q}_i {\cal Q}_j ,
\ee
with the following nonvanishing Poisson brackets:
\ba\label{11}
\{X_i ,X_j \} &=&\gamma \varepsilon_{ij},\;\;\; \{X_i ,P_j \}=\delta_{ij},\nonumber\\
\{{\cal Q}_i ,{\cal Q}_j \}&=& -\gamma\varepsilon_{ij}.
\ea

We can see that our Lagrangian is now separated  into two disconnected parts describing the ``external'' and ``internal'' degrees of freedom in a noncommutative phase space, parametrized by the variables $(X_i ,P_i)$(external structure) and ${\cal Q}_i$(internal structure) \cite{Plyushchay1}. 

Now we shall introduce an interaction to the ``external'' sector, equation (\ref{09})(which do not modify the internal sector), represented by a potential energy term $U(X)$ involving noncommutative variables, as follows
\be\label{11.0}
L_{\rm ext}=P_i \dot X_i + {\gamma\over2}\varepsilon_{ij}P_i \dot P_j -{1\over 2m}g_{ij}P_i P_j -U(X)\,\,.
\ee
This leads to a deformation of the constraint algebra, since the constraint now involves a derivative of the potential \cite{Albert1}.

\section{The supersymmetric model in $N=2$}

To obtain the supersymmetric extension of the model described by the Lagrangian (\ref{11.0}), for each space commuting coordinate, representing the degrees of freedom of the system, we associate one anticommuting variable, which are the well known Grassmannian variables. We are considering only the $N=2$ SUSY for a non-relativistic particle, which is described by the intoduction of two real Grassmannian variables $\Theta$ and $\bar\Theta$ (the Hermitian conjugate of $\Theta$) in the configuration space, but all the dynamics are represented by the time $t$ \cite{Galvão,Junker}.

Furthermore, intoducing the Taylor expansion for the real scalar supercoordinate as
\be
\label{11.1}
X_i \rightarrow {\cal X}_i(t,\Theta,\bar\Theta) =X_i(t) + i\psi_i(t)\Theta +i\bar\Theta \bar{\psi}_i(t) +\bar{\Theta}\Theta F_i(t)
\ee
and their canonical supermomenta
\be\label{11.2}
P_i(t) \rightarrow {\cal P}_i(t,\Theta,\bar\Theta)=i\eta_i(t)-i\Theta\left(P_i(t)+if_i(t)\right)-\bar{\Theta}\Theta\dot\eta_i(t),
\ee
which under the infinitesimal supersymmetry transformation law
\ba\label{11.2.0}
\delta t &=& (\epsilon Q + \bar{\epsilon}\bar{Q})t,\nonumber\\
\delta \Theta &=& (\epsilon Q)\Theta,\\
\delta \bar{\Theta} &=& (\bar{\epsilon}\bar{Q})\bar{\Theta},\nonumber
\ea
furnish
\ba\label{11.2.1}
\delta {\cal X}_i &=&(\epsilon\bar{Q} + \bar{\epsilon}Q){\cal X}_i \\
\delta{\cal P}_i &=& (\epsilon\bar{Q} + \bar{\epsilon}Q){\cal P}_i\,\,,
\ea
where $Q$ and $\bar{Q}$ are the two SUSY generators
\be\label{11.2.3}
Q=\frac{\partial}{\partial \bar{\Theta}}+i\Theta\frac{\partial}{\partial t},\;\;\;\;\; \bar{Q}=\frac{\partial}{\partial {\Theta}}+i\bar{\Theta}\frac{\partial}{\partial t}.
\ee

In terms of $(X_i(t), P_i(t), F_i,f_i)$ bosonic (even) components and $(\psi_i(t),\bar{\psi}_i(t),\eta_i(t))$ fermionic (odd) components, we get the following transformations,
\ba\label{11.2.4}
\delta X_i &=& i(\bar{\epsilon}\bar{\psi}_i -\epsilon\psi_i )\nonumber\\
\delta \psi_i &=& -\bar{\epsilon}(\dot{X}_i -iF_i )\nonumber\\
\delta \bar{\psi}_i &=& -\epsilon (\dot{X}_i +iF_i )\\
\delta F_i &=& \epsilon\dot{\psi}_i + \bar{\epsilon}\dot{\bar{\psi}}_i ,\nonumber
\ea
and
\ba\label{11.2.5}
\delta\eta_i &=& -\epsilon(P_i +if_i)\nonumber\\
\delta P_i &=& 0 \nonumber\\
\delta f_i &=& 2\bar{\epsilon}\dot{\eta}_i\\
\delta \dot{\eta}_i &=& -\epsilon(\dot{P}_i + i\dot{f}_i ).\nonumber
\ea 
Notice that the supersymmetry mixes the even and odd coordinates. Carrying out a variation in the even components we obtain the odd components and vice-versa.

The super-Lagrangian for the superpoint particle with $N=2$, invariant under the transformations (\ref{11.2.4}) and (\ref{11.2.5}),  can be written as the following integral (we use for simplicity that $m=1$):
\ba\label{11.3}
\bar{L}_{\rm ext}&=&\frac{1}{2}\int{\rm d}\Theta {\rm d}{\bar\Theta}\left[ \frac{}{}\(\bar{D}{\cal X}_i \bar{\cal P}_i+{\cal P}_iD{\cal X}_i\)\right. \nonumber\\ &+&\left. \frac{\gamma}{2}\varepsilon_{ij}\({\cal P}_i\dot{\bar{\cal P}}_j +\dot{\cal P}_j \bar{\cal P}_i \) -\frac{1}{2}g_{ij}\({\cal P}_i\bar{\cal P}_j +{\cal P}_j\bar{\cal P}_i \)\right]\nonumber\\
&-& \int{\rm d}\Theta {\rm d}{\bar\Theta}U[{\cal X}(t,\Theta,\bar\Theta )]
\ea
where $D$ is the covariant derivative $(D=\partial_{\Theta}-i\bar\Theta\partial_t)$ and $\bar D$ is its Hermitian conjugate. The $U[{\cal X}]$ is a polynomial function of the supercoordinate

Expanding  the superpotential $U[{\cal X}]$ in Taylor series and maintaining $\Theta\bar\Theta$ (because only these terms remain after  integrations on Grassmannian  variables $\Theta$ and $\bar\Theta$), we have that
\ba\label{11.4}
U[{\cal X}]&=&{\cal X}_i \frac{\partial U[X(t)]}{\partial X_i} + \frac{{\cal X}_i{\cal X}_j^*}{2}\frac{\partial^2 U[X(t)]}{\partial X_i \partial X_j}+ ...\\
&=&F_i \bar\Theta \Theta \partial_i U[ X(t)]+\bar\Theta \Theta \psi_i \bar{\psi}_j \partial_i \partial_j U[ X (t)]+...\nonumber
\ea
where the derivatives $\partial_i =\frac{\partial}{\partial X_i}$ are such that $\Theta=0=\bar\Theta$, which are functions only of the $X(t)$ even coordinate. Substituting euqation (\ref{11.4}) in equation (\ref{11.3}), we  obtain after integrations 
\ba\label{11.5}
\bar{L}_{\rm ext}&=& L^{(0)}_{\rm ext} -\frac{1}{2}g_{ij}f_i f_j-F_i f_i +\frac{\gamma}{2}\varepsilon_{ij}f_i\dot{f}_j \nonumber\\ &-&i\(\bar{\psi}_i \dot{\bar \eta}_i  -\dot{\eta}_i\psi_i \) -ig_{ij}\dot{\eta}_i \bar{\eta}_j + i\gamma \varepsilon_{ij}\dot{\eta}_i\dot{\bar\eta}_j \nonumber\\&-&F_i \partial_i U[X(t)]-\psi_i \bar{\psi}_j \partial_i \partial_j U[X (t)],
\ea
which is the complete Lagrangian for $N=2$. 

The bosonic component $F_i$ is not a  dynamic variable. In this case, using the Euler-Lagrange equations for the auxiliary variables $f_i$ and $F_i$, we obtain:
\ba
f_i(t)&=& \partial_i U[X(t)],\label{11.6}\\
F_i(t)&=& g_{ij}f_j -\gamma\varepsilon_{ij}\dot f_j \nonumber\\
&=&g_{ij} \partial_j U[X(t)]-\gamma\varepsilon_{ij} \partial_j\partial_k U[X] \dot X_k (t),\label{11.7}
\ea
where we have to eliminate the variable $f_i$ as well as its derivative in $F_i$.  Now, substituting the (\ref{11.6}) and (\ref{11.7}) in  (\ref{11.5}) the auxiliary variables can be completely eliminated, so that 
\ba\label{11.8}
\bar{L}_{(N=2)\rm{ext}}&=&L_{\rm{ext}}^{(0)} -\frac{1}{2}g_{ij}\partial_i U\partial_j U+\frac{\gamma}{2}\varepsilon_{ij}\partial_i U\partial_j\partial_k U\dot X_k \nonumber \\&-&i\(\bar{\psi}_i \dot{\bar \eta}_i  -\dot{\eta}_i\psi_i \) -ig_{ij}\dot{\eta}_i \bar{\eta}_j + i\gamma \varepsilon_{ij}\dot{\eta}_i\dot{\bar\eta}_j\nonumber\\ &-&\psi_i\bar{\psi}_j \partial_i\partial_j U,
\ea

Note that, as in \cite{Lukierski2}, we can rewrite equation (\ref{11.8})
 as 
\ba\label{11.9}
\bar{L}_{(N=2)\rm{ext}}&=&{L}_{\rm{ext}}^{(0)} +A_k(X,t)\dot X_k +A_0(X,t)+ \nonumber \\&-&i\(\bar{\psi}_i \dot{\bar \eta}_i  -\dot{\eta}_i\psi_i \) -ig_{ij}\dot{\eta}_i \bar{\eta}_j + i\gamma \varepsilon_{ij}\dot{\eta}_i\dot{\bar\eta}_j\nonumber\\ &-&\psi_i\bar{\psi}_j\partial_i\partial_j U,
\ea
that is invariant under standard gauge transformations $A_{\mu}\rightarrow A_{\mu}^{\prime}=A_{\mu} +\partial_{\mu}\Lambda$, where
\be\label{11.10}
A_0( X,t) =-\frac{1}{2}g_{ij}\partial_i U\partial_j U
\ee
and 
\be\label{11.12}
A_k(X,t)=\frac{\gamma}{2}\varepsilon_{ij}\partial_i U\partial_j\partial_k U.
\ee
were identifided in \cite{Lukierski2} with the scalar potential $A_0$ (that in this case have a pseudo-Euclidean metric) and the vector potential $A_k$.  The vector potential introduce a magnetic field $B=\varepsilon_{ij}\partial_i A_j$ given by
\be\label{11.13}
B( X)=\frac{\gamma}{2}\varepsilon_{ik}\varepsilon_{lj}\left(\partial_i\partial_l U\right)\left(\partial_j\partial_k U\right)
\ee
where we see that the noncommutativity introduced by the parameter $\gamma$ generates a constant magnetic field \cite{Lukierski2}.

The Euler-Lagrange equations, in this case, are
\ba\label{11.14}
m^* \dot X_i &=& g_{ij}P_j -me\gamma\varepsilon_{ij}E_j + m\gamma\varepsilon_{ij}\psi_l\bar{\psi}_k\partial_l\partial_k\partial_j U,\nonumber\\
\dot P_i &=& eB\varepsilon_{ij}\dot X_j + eE_i - \psi_l\bar{\psi}_j\partial_l\partial_j\partial_i U,
\ea
where $E_i$ and $B$ are the electric and magnetic field, respectively, and $m^{*} =m(1-e\gamma B)$ is an effective mass. But, such a way of introducing electromagnetic interaction modifies the symplectic structure of the system which determines the noncommutative phase-space geometry, for the bosonic sector, equation (\ref{11}),
\ba\label{11.15}
\{X_i,X_j \}&=&\frac{m}{m^*}\gamma\varepsilon_{ij},\;\; \{X_i ,P_j \}=\frac{m}{m^*}\delta_{ij},\nonumber\\
\{P_i ,P_j \}&=&\frac{m}{m^*}eB\varepsilon_{ij},
\ea
which implies an analysis of the value $e\gamma B \neq 1$ in order to avoid a singularity \cite{Plyushchay2,Duval}. To the fermionic sector, the Euler-Lagrange equations are
\ba\label{11.16}
i\gamma\varepsilon_{ij}{\ddot{\bar\eta}}_j -ig_{ij}{\dot{\bar\eta}}_j +i\dot\psi &=&0,\nonumber\\
-i\gamma\varepsilon_{ij}{\ddot{\eta}}_j -ig_{ij}{\dot\eta}_j -i\dot{\bar\psi} &=&0,
\ea
for the fermionic varialbles $(\eta,\bar\eta)$. For the fermionic variables  $(\psi_i ,\bar{\psi}_i )$ the Euler-Lagrange equations are
\ba\label{11.17}
i\dot\eta_i + \bar\psi_j \partial_i\partial_j U &=& 0\nonumber\\
i\dot{\bar\eta}_i - \psi_l \partial_i \partial_j U &=&0\,\,.
\ea
where the fermionic variables $(\psi_i, \bar{\psi}_i )$ do not have dynamics.

The canonical Hamiltonian  for the $N=2$ SUSY is given by
\ba\label{11.18}
\bar{H}&=&\dot{X}_i\frac{\partial\bar{L}}{\partial \dot{X}_i}+\frac{\partial\bar{L}}{\partial\dot{\psi}_i}
\dot{\psi}_i+\frac{\partial\bar{L}}{\partial\dot{\bar\psi}_i}
\dot{\bar\psi}_i+\frac{\partial\bar{L}}{\partial\dot{\eta}_i}
\dot{\eta}_i+\frac{\partial\bar{L}}{\partial\dot{\bar\eta}_i}
\dot{\bar\eta}_i - \bar{L}_{(N=2)}\nonumber\\
&=& \frac{1}{2m}g_{ij}P_i P_j -A_{0} +\psi_i \bar{\psi}_j \partial_i \partial_j U(X),
\ea
which provides a mixed potential term, with a dynamical variable of the particle $A_0$ and the Grassmannian  variables $(\psi_i ,\bar{\psi}_i )$.

There is one other way to introduce the minimal electromagnetic interaction.  
It is through the transformation $P_i \rightarrow {\cal P}_i =P_i + eA_i (X_i ,t)$ in the Hamiltonian, that preserve the symplectic structure of equation (\ref{11}). In \cite{Lukierski2} this transformation has been considered and it leads to the same expression for the magnetic field Eq.(\ref{11.13}).

\section{Remarks and conclusions}

A fundamental property of all charged particles is that the electromagnetic energy is radiated whenever they are accelerated.  The recoil momentum of the photons emitted during this process is equivalent to a reaction force corresponding to the self-interaction of the particle with its own  electromagnetic field, which originates radiation damping.

The process of RD is important in many areas of electron accelerator operation, like in recent experiments with intense laser relativistic electron scattering at laser frequencies and field strengths where radiation reaction forces begin to become significant.
 
In \cite{Albert1} some of us introduced an alternative approach to canonical quantization of the RD based on doubling the degrees of freedom.  A Lagrangian model for the system with a Chern-Simons-like term with high order derivative was obtained.  In \cite{Albert2}  it was introduced the $N=1$ supersymmetric version of the RD in the Grassman superspace and the $N=2$ version was constructed in \cite{Lukierski2}.

Here the supersymmetric model was split into ``external" and ``internal" degrees of freedom of the supersymmetric model in terms of new variables, where the RD constant introduced noncommutativity in the coordinate sector.
We presented a way to introduce an electromagnetic coupling.  

We carried out the supersymmetric $N=2$ extension of the RD model and realized that the noncommutativity introduced by the parameter generates a constant magnetic field.

Also in this work, we used an alternative way to construct a dual equivalent action to the RD one, a dualization procedure.  It used the Noether technique which is independent of dimensions and imposes a gauge symmetry which is believed to be hidden in the theory.  The main ingredient is an auxiliary field which is eliminated through the equations of motion and the final action is an effective one depending only on this original variables.

\section{ Acknowledgments}
This work is partially supported by FAPERJ, FAPEMIG and CNPq, Brazilian Research Agencies.


\begin{thebibliography}{99}

\bibitem{uz}  W. G. Unruh and W. H. Zurek, Phys. Rev. D 40 (1961) 37.

\bibitem{cl}  A. O. Caldeira and A. J. Legget, Phys. Rev. Lett. 46 (1981) 211.

\bibitem{whb}  W. S. Warren, S. L. Hammes and J. L. Bates, J. Chem. Phys. 91 (1989) 5895.

\bibitem{mt}  A. C. R. Mendes and F. I. Takakura, Phys. Rev. E 64 (2001) 056501.

\bibitem{bc}   P. M. V. B. Barone and A. O. Caldeira, Phys. Rev. A 43 (1991) 57.

\bibitem{fv}  R. P. Feynman and F. L. Vernon Jr., Ann. Phys. 24 (1963) 118.

\bibitem{cl2}  A. O. Caldeira and A. J. Legget, Ann. Phys. 121 (1983) 587.

\bibitem{cl3}  A. O. Caldeira and A. J. Legget, Physica A 149 (1983) 374.

\bibitem{bm}  R. Banerjee and P. Mukherjee, quant-ph/0205143; ibid, quant-ph/0108055.

\bibitem{ft}  H. Feshbach and Y. Tikochinsky, Trans. N. Y. Acad. Sci., Ser II 38 (1977) 44.

\bibitem{hjp}  W. Heitler, ``The quantum theory of radiation", 3nd ed., Dover, 1970; J.D. Jackson, ``Classical electrodynamics", 2nd ed., 
Chaps. 14 and 17, Wiley, New York 1975; G. N. Plass, Rev. Mod. Phys. 33 (1961) 37. 

\bibitem{becker}  R. Becker, ``Electromagnetic Fields and Interactions", vol. 1 and 2, Blaidsell, New York, 1965.

\bibitem{lorentz}  H. A. Lorentz, ``The theory of electrons", second ed., Dover, New York, 1952.

\bibitem{walker}  R. P. Walker, ``Radiation damping", in: Proceeding, General Accelerator Physics, CERN 
Genebra.

\bibitem{hk}  F. V. Hartemann and A. K. Kerman, Phys. Rev. Lett. 76 (1996) 624.

\bibitem{bula}  C. Bula et al, Phys. Rev. Lett. 76 (1996) 3116.

\bibitem{mendes}  A. C. R. Mendes and P. M. V. B. Barone, ``Phenomenological and microscopic approach of the radiative radiation damping", unpublished monograph (in Portuguese), UFJF, 1995.

\bibitem{Barone}   P. M. V. B. Barone and A. C. R. Mendes, Phys. Lett. A 364 (2007) 438.

\bibitem{Plyushchay1}   P. A. Horv\'athy and M. S. Plyushchay, JHEP 0206 (2002) 033; Phys. Lett. B 595 (2004) 547.

\bibitem{Lukierski}   J. Lukierski, P. C. Stichel and W. J. Zakrzewski, Ann. Phys. 260 (1997) 224. 

\bibitem{ls}   J. Lopuszanski and P. C. Stichel, Fortschr. Phys. 45 (1997) 79.

\bibitem{Lukierski2}  J. Lukierski, P. C. Stichel and W. J. Zakrzewski, Phys. Lett. B 602 (2004) 249.


\bibitem{Albert1}   A. C. R. Mendes, F. I. Takakura, C. Neves and W. Oliveira, Eur. Phys. J. C 45 (2006) 257.


\bibitem{Plyushchay}  M. S. Plyushchay, Nucl. Phys B 362 (1991) 54; Nucl. Phys. B 262 (1991) 71.


\bibitem{Albert2}  A. C. R. Mendes, F. I. Takakura, C. Neves and W. Oliveira, J. Phys. A 38 (2005) 9387.

\bibitem{coleman}  S. Coleman, Phys. Rev. D 11(1975) 2088; S. Mandelstan, Phys. Rev. D 11 (1975) 3026.

\bibitem{dns}  P. H. Damgaard, H. B. Nielsen and R. Scollacher, Nucl. Phys. B 385 (1992) 227; Nucl. Phys. B 414 (1994) 541; Phys. Lett. B 296 (1992) 132; Phys. Lett. B 332 (1994) 131; P. H. Damgaard and H. B. Nielsen, Nucl. Phys. B 433 (1995) 671.

\bibitem{luscher}  M. L\"uscher, Nucl. Phys. B 326 (1989) 557.  

\bibitem{marino}  E. C. Marino, Phys. Lett. B 263 (1991) 63.

\bibitem{iw}  A. Ilha and C. Wotzasek, Nucl. Phys. B 604 (2001) 426.

\bibitem{abw}  E. M. C. Abreu, R. Banerjee and C. Wotzasek, Nucl. Phys. B 509 (1998) 519; E. M. C. Abreu, R. Menezes and C. Wotzasek, Phys. Rev. D 71 (2005) 065004.

\bibitem{bw}  R. Banerjee and C. Wotzasek, Nucl. Phys. 527 (1998) 402.

\bibitem{iw2}  A. Ilha and C. Wotzasek, Phys. Lett. B 519 (2001) 169.

\bibitem{ainrw}  M. A. Anacleto, A. Ilha, J. R. S. Nascimento, R. F. Ribeiro and C. Wotzasek, Phys. Lett. 504 (2001) 268.

\bibitem{binrw}  D. Bazeia, A. Ilha, J. R. S. Nascimento, R. F. Ribeiro and C. Wotzasek, Phys. Lett. B 510 (2001) 329.

\bibitem{vyt}  A. S. Vytheeswaran, ``Hidden symmetries in second class constrained systems: are new fields necessary", hep-th/0007230.

\bibitem{nw}  C. Neves and C. Wotzasek, J. Phys. A 33 (2000) 6447; A. C. R. Mendes, W.
Oliveira, C. Neves and D. C. Rodrigues; Nucl. Phys. B (Proc. Suppl.) 127 (2004), 170;
E. M. C. Abreu, A. Ilha, C. Neves and C. Wotzasek, Phys. Rev. D 61 (2000) 025014.

\bibitem{Faddeev}  L. D. Faddeev and R. Jackiw, Phys. Rev. Lett. 60 (1988) 1692.

\bibitem{Galvão} C. A. P. Galv\~ao and C. Teitelboim, J. Math. Phys. {\bf 21} (1980) 1863.

\bibitem{Junker} G. Junker and S. Mattiensen, J. Phys. A {\bf 27} (1995) L751. 

\bibitem{Plyushchay2}  P. A. Horv\'athy and M. S. Plyushchay, Nucl. Phys. B 714 (2005) 269, hep-th/0502040.

\bibitem{Duval}  C. Duval and P. A. Horv\'athy, J. Phys. A 34 (2001) 10097.

\bibitem{Seiberg}  N. Seiberg and E. Witten, JHEP 9909 (1999) 032. 

\end{thebibliography}
\end{document}